\pdfoutput=1

\documentclass[reprint,longbibliography,amsmath,amssymb,aps,prb]{revtex4-2}

\usepackage{graphicx}
\usepackage{dcolumn}
\usepackage{bm}
\usepackage[colorlinks=true, linkcolor=blue, citecolor=red]{hyperref}

\def\ths{\theta^\text{S}} 
\def\thv{\theta^\text{V}} 

\def\us{u^\text{S}}

\def\a{\alpha}
\def\b{\beta}
\def\nab{\mbox{\boldmath $\nabla$}}

\def\rn{{\rho_\text{n}}}
\def\rs{{\rho_\text{s}}}

\def\bvs{{\bf v}_\text{s}}
\def\bvn{{\bf v}_\text{n}}
\def\vs{{v_\text{s}}}
\def\vn{{v_\text{n}}}

\def\half{\frac{1}{2}}
\def\lag{\mathcal{L}}
\def\d{\delta}
\def\p{\partial}

\begin{document}
	
\title{Vortex dynamics in two-dimensional supersolids}

\author{Chi-Deuk Yoo}
\affiliation{Century College, White Bear Lake, Minnesota 55110}

\author{Alan T. Dorsey}
\affiliation{Department of Physics and Astronomy, University of Georgia, Athens, Georgia 30602}

\date{September 7, 2024}

\begin{abstract}
We investigate the dynamics of quantized vortices in a model two-dimensional supersolid. Starting from an effective action that captures the dynamics of the superfluid condensate and its coupling to the lattice displacements, we integrate out the low-energy Goldstone modes--the phonons of the solid and the superfluid condensate--to arrive at an effective action for the vortices in the condensate. In the low-velocity limit we calculate the effective inertial mass for the vortices, and we find that the mass has a logarithmic frequency dependence, similar to the inertial mass found in superfluid vortices. The vortex dynamics also includes a Magnus force term in the equation of motion that arises from the Berry phase in the effective action.
\end{abstract}

\maketitle

\section{Introduction}

Supersolidity is an elusive phase of matter with coexisting off-diagonal long-range order \cite{Yang1962a} and crystalline order (with the accompanying shear rigidity of a solid). The foundational work on Bose condensation by Penrose and Onsager \cite{Penrose1951a,Penrose1956a} ruled out such a phase for commensurate Bose solids (i.e., crystals with exactly one boson per lattice site). However, Andreev and Lifshitz \cite{Andreev1969a} and Chester \cite{Chester1970a} noted that a Bose crystal with defects, in the form of vacancies, might support a supersolid phase; in the Andreev and Lifshitz description, the dilute vacancies form a superfluid that suffuses the crystal, and leads to novel collective modes which couple the superfluidity to the elasticity of the solid. Leggett \cite{Leggett1970a,Leggett1998a} suggested that such a phase could be detected by observing nonclassical rotational inertia (NCRI) in a rotating supersolid. Two decades of experimental searches failed to yield a definitive observation of this phase \cite{Meisel1992a}; after a long dormant period, in 2004 Kim and Chan announced the observation of a small NCRI signal in crystals of $^4$He \cite{Kim2004a,Kim2004b}. This observation ushered in a decade of intense experimental and theoretical work to elucidate the properties of the supersolid phase of matter; we refer the reader to Ref.~\cite{Boninsegni2012a} for a summary. Alas, successively refined experimental techniques resulted in smaller observed values of NCRI, until in 2012 the NCRI effect was shown to be absent in carefully prepared $^4$He crystals \cite{Kim2012a}. 

In parallel with the work on $^4$He, the ultracold-atom community developed techniques to trap and condense dipolar bosons, including lanthanides such as $^{164}$Dy and $^{166}$Er. Like their classical brethren, these quantum dipolar systems organize into complex patterns, including labyrinths, stripes, and bubbles; for accessible reviews, see \cite{Langen2022a,Recati2023a}. At sufficiently high densities, these systems not only break translational symmetry but also become phase coherent, producing a dipolar supersolid \cite{Tanzi2019a,Tanzi2019b,Bottcher2019a,Chomaz2019a,Natale2019a}. These new experimental studies motivate the present work on topological defects in supersolids. 

Topological defects are ubiquitous in systems with a broken continuous symmetry, be they vortices in a superfluid or dislocations in a solid. In a supersolid, we have both types of defect; barring a hidden symmetry or fine-tuning of parameters, we expect these defects to interact. These point defects are particularly important in two-dimensions at finite temperatures, as their proliferation destroys the ordered phase (vortices destroying the broken $U(1)$ symmetry of the superfluid and dislocations the broken translational symmetry of the solid). In this work we focus on vortices in a two-dimensional supersolid, leaving to future work the discussion of dislocations and their interaction with vortices. We note that the observation of vortices in rotating supersolid $^{164}$Dy has recently been reported in Ref.~\cite{Casotti2024a}. 

The remainder of this paper is organized as follows. We use a hydrodynamic approach, which captures the physics of long-lived, low frequency excitations that emerge from conserved densities and the Goldstone modes of the broken symmetry phases. The hydrodynamic approach to supersolids originated with Andreev and Lifshitz \cite{Andreev1969a}, with subsequent developments \cite{Saslow1977a,Liu1978a,Pomeau1994a,Stoof1996a,Dorsey2006a,Sears2010a,Heinonen2019a,Hofmann2021a}. The hydrodynamic equations can be derived from a Lagrangian and an action principle \cite{Son2005a,Ye2008a,Josserand2007a,Peletminskii2009a,Yoo2010a,YooThesis2009,During2011a,Rakic2024a}. The Lagrangian formulation is particularly convenient for developing the dynamics for vortices in supersolids, and we will use the formulation developed by Yoo and Dorsey \cite{Yoo2010a,YooThesis2009}, which is summarized in the appendix. We then introduce the vortex degrees of freedom, and integrate out the smoothly varying fields; due to the compressible low-energy superfluid and elastic excitations, the resulting vortex-vortex interaction is nonlocal in space and time. In the low-velocity limit, this nonlocal interaction produces an effective inertial mass with a logarithmic frequency dependence. We also make explicit the Magnus force, which results from the Berry's phase term in the effective action \cite{Ao1993a,Zhu1996a,Thouless1996a}. 

\section{Effective action for supersolids}

We start with the quadratic action for isotropic supersolids, derived in Ref.~\cite{Yoo2010a} 
(see Appendix~A for its derivation and a brief summary of Ref.~\cite{Yoo2010a}), expressed in terms of the dynamical variables which are the superfluid phase $\theta$ and components of the lattice displacement $u_i$: 
\begin{equation}
\begin{split}
S[\theta, u_i] = & \frac{1}{2} \int d t \int d^2 x \bigg[ 
 - 2 \rho \partial_t \theta - 2 \lambda_{ij} u_{ij}
+ \rho^2 \chi (\partial_t \theta)^2 
\\
&
- \rho_\text{s} (\partial_i \theta)^2 + \rho_\text{n} (\partial_t u_i)^2 
- (\tilde{\lambda} - \rho^2 \gamma^2 \chi) u_{ii}^2
\\
& 
- 2 \tilde{\mu} u_{ij}^2  
- ( \rho_\text{n} - \rho^2 \gamma \chi) ( \partial_t u_i \partial_i \theta + u_{ii} \partial_t \theta ) \bigg],
\label{SS.action1}
\end{split}
\end{equation}
with $\rho_\text{n}$ the density of normal component, $\rho_\text{s}$ the density of the super-component, and $\rho=\rho_\text{n}+\rho_\text{s}$ the total density; $\tilde{\lambda}$ and $\tilde{\mu}$ are the Lam\'e coefficients \cite{Landau.elasticity.70} for an isotropic solid; $\gamma$ is a phenomenological coupling constant between the strain and the density; $\chi^{-1}$ is the isothermal compressibility; and
\begin{equation}
u_{ij} = \frac{1}{2}\left(\partial_i u_j + \partial_j u_i\right)
\end{equation}
is the linearized symmetric strain tensor  ($\partial_i\equiv \partial/\partial x_i$ with Roman subscripts reserved for Cartesian coordinates, and $\partial_t \equiv \partial/\partial t$ in what follows). A few comments on the above action are helpful here. 

First, the density fluctuation was integrated out of the action, renormalizing the coupling between the normal velocity ($\partial_t u_i$) and the superfluid velocity ($\partial_i \theta$) from $\rho_\text{n}$ to $ \rho_\text{n} - \rho^2 \gamma \chi$, and the elastic constant from $\tilde{\lambda}$ to $\tilde{\lambda} - \rho^2 \gamma^2 \chi$. 
In order to keep the notation simple, we use $\tilde{\Lambda} \equiv \tilde{\lambda} - \rho^2 \gamma^2 \chi$ and $\tilde{\Gamma} \equiv \rho_\text{n} - \rho^2 \gamma \chi$ hereafter. Note that when $\tilde{\Gamma}=0~(\rho_\text{n} = \rho^2 \gamma \chi)$, the elastic part and superfluid terms become completely decoupled.

Second, we have explicitly isolated the interaction terms between superfluidity and elasticity as $\partial_t u_i \partial_i \theta$ and $u_{ii} \partial_t \theta$. Note that the phase $\theta$ is \textit{odd} under time reversal (as gradients of $\theta$ produce the superfluid velocity) and the $u_i$ are \textit{even} under time reversal, so the two coupling terms above are both \textit{even} under time reversal as required for a Lagrangian that is symmetric under time reversal. If the fields $u_i$ and $\theta$ are analytic, then an integration by parts allows one to combine these two terms. However, this is no longer possible with the singular fields due to topological defects. 
             
Third, the first two terms in the action, which are linear in $\theta$ and $u_i$, generally can be ignored as long as vortices and dislocations are absent because they are the total derivatives of analytical dynamical variables. However, the first term is responsible for the Magnus force on vortices \cite{Ao1993a,Zhu1996a,Thouless1996a} and the second term produces the Peach-Koehler force on a dislocation \cite{Peach1950a}. In this work, we disregard the second term as we focus only on vortices--dislocations and their coupling with vortices in a supersolid will be presented elsewhere. 

The equations of motion can be obtained by taking the variations of $S$ with respect to $\theta$ and $u_i$,
\begin{equation}
\tilde{\chi} \partial_t^2 \theta - \rho_\text{s} \partial^2 \theta - \frac{\tilde{\Gamma}}{2} (\partial_i \partial_t u_i + \partial_t u_{ii}) = 0,
\label{eqn.motion1}
\end{equation}
\begin{equation}
\rho_\text{n} \partial_t^2 u_i - \partial_j \sigma_{ij} - \frac{\tilde{\Gamma}}{2} (\partial_t \partial_i \theta + \partial_i \partial_t \theta) = 0,
\label{eqn.motion2}
\end{equation}
where $\tilde{\chi} \equiv \rho^2 \chi$, and the elastic stress tensor
\begin{equation}
\sigma_{ij} = \tilde{\Lambda} \delta_{ij} u_{kk} + 2\tilde{\mu} u_{ij}.
\end{equation}
Equations (\ref{eqn.motion1}) and (\ref{eqn.motion2}) are three equations of motion that yield three pairs of sound modes for two-dimensional isotropic supersolids: transverse (speed $c_T$) and longitudinal (speed $c_L$) elastic modes and a longitudinal second sound mode (speed $c_2$); expressions for the sound speeds can be found in Ref.~\cite{Yoo2010a}, Eqs.~(48)-(51). The above Lagrangian was also used recently in Ref.~\cite{Platt2024a} to study the collective modes of a one-dimensional supersolid phase. 

In what follows we take vortices into account by introducing singular parts of $\theta$ in the action, Eq.~(\ref{SS.action1}), and integrate out the smooth superfluid and elastic fields, resulting in an effective action for the vortices.

\section{Vortices in two-dimensional supersolids}

For a vortex $\alpha$ of integer winding number (or charge) $e^\alpha$ the counterclockwise line integral of $\theta$ is given by
\begin{equation}
\oint d\theta = \frac{h}{m} e^\alpha,
\label{lineintegral.1}
\end{equation}
where $h$ is the Planck constant and $m$ the particle mass (Greek superscripts label vortices). Separating $\theta$ into its smooth part $\ths$ and singular (or vortex) part $\thv$ as
\begin{equation}
\theta = \ths + \thv,
\label{theta.total}
\end{equation}
we see the velocity of the super-component can be written as the sum of a longitudinal part $\partial_i \ths$ and a transverse part $\partial_i \thv$. Taking Eq.~(\ref{lineintegral.1}) along a path enclosing $n_\text{V}$ vortices,
\begin{equation}
\oint d \theta = \oint d \thv = \frac{h}{m} \sum_{\a = 1}^{n_\text{V}} e^\a,
\label{lineintegral.2}
\end{equation}
with the $\a$-th vortex located at ${\bf x}^\a(t) = [x^\a(t), y^\a(t)]$, we have
\begin{equation}
\nab \times \nab \thv = \frac{h}{m} \hat{z} \sum_\a e^\a \delta^{(2)}({\bf x} - {\bf x}^\a).
\label{curl.of.grad.theta}
\end{equation}

Next, we substitute Eq.~(\ref{theta.total}) into the equations of motion, Eqs.~(\ref{eqn.motion1}) and (\ref{eqn.motion2}):
\begin{equation}
\tilde{\chi} \partial_t^2 \ths - \rho_\text{s} \partial^2 \ths - \tilde{\Gamma} \partial_t \us_{ii} = 
-\tilde{\chi} \partial_t^2 \thv,
\label{eqn.motion3}
\end{equation}
\begin{equation}
\begin{split}
\rho_\text{n} \partial_t^2 u_i -& \partial_j \sigma_{ij} - \tilde{\Gamma} \partial_i \partial_t \ths = 
\\
&
+\frac{1}{2}\tilde{\Gamma} \partial_t (\partial_i \thv) 
+\frac{1}{2}\tilde{\Gamma} \partial_i (\partial_t \thv).
\label{eqn.motion4}
\end{split}
\end{equation}
Inverting these inhomogeneous differential equations and eliminating $u_i$ and $\ths$ from the action Eq.~(\ref{SS.action1}), one obtains
%
\begin{equation}
S[\thv] = S^\text{V}_{1}[\thv] + S^\text{V}_{2}[\thv], 
\end{equation}
where
\begin{equation}
S^\text{V}_{1}[\thv] = - \rho \int dt \int d^2 x \; \partial_t \thv, 
\end{equation}
\begin{widetext}
\begin{eqnarray}
S^\text{V}_{2}[\thv] &=& 
- \frac{1}{2} \int \frac{d \omega}{2\pi} \int \frac{d^2 q}{(2\pi)^2}
\Bigg[ \frac{q^2}{\Delta} 
\bigg( \rs\rn + \frac{\tilde{\Gamma}^2}{4} \bigg) \omega^2 - \rs \bigg( \tilde{\Lambda} + 2 \tilde{\mu} + \frac{\tilde{\Gamma}^2}{4\tilde{\chi}} \bigg) q^2 \Bigg] 
(\partial_t \thv)({\bf q},t) (\partial_t \thv)(-{\bf q},-t)
\nonumber
\\
&&
- \frac{1}{2} \int \frac{d \omega}{2\pi} \int \frac{d^2 q}{(2\pi)^2}
\bigg( \rho_\text{s} + \frac{\tilde{\Gamma}^2}{4} \frac{\omega^2}{\rn \omega^2 - \tilde{\mu} q^2} \bigg) 
(\partial_i \thv)({\bf q},t) (\partial_i \thv)(-{\bf q},-t),
\label{SS.action.V2}
\end{eqnarray}
\end{widetext}
where 
\begin{eqnarray}
\Delta &=& \rho_\text{n} \omega^4  
- \left[ \tilde{\Lambda} + 2 \tilde{\mu} + \frac{\rn \rs}{\tilde{\chi}} + \frac{\tilde{\Gamma}^2}{{\tilde{\chi}}} \right] q^2 \omega^2 
\nonumber
\\
&&
+ \frac{\rho_\text{s}}{\tilde{\chi}} \left(\tilde{\Lambda} + 2 \tilde{\mu}\right)q^4.
\label{Delta.determinant}
\end{eqnarray}
Note that in the action $S_2^\text{V}$ there is no coupling between $\partial_i \thv$ and $\partial_t \thv$. 

\section{Vortex dynamics}

\subsection{Effective Action for Vortices}

Now we turn to the dynamical properties of vortices by deriving an effective action in terms of their coordinates. We follow the general development for superfluid vortices found in Refs.~\cite{Eckern1989a, Arovas1997a}. Both of these works find that a coupling of vortex motion to the low-energy excitations, superfluid phonons, leads to a logarithmic frequency-dependent effective mass for the vortices. Analogously, we will find in a supersolid the interactions are mediated by the superfluid phonons as well as the longitudinal and transverse elastic modes, with concomitant contributions to the effect mass. 

To capture the singular nature of the phase variable specified in Eq.~(\ref{curl.of.grad.theta}) we take 
\begin{equation}
\thv =
\frac{\hbar}{m} \sum_\a e^\a \arctan \left[ \frac{y-y^\a(t)}{x-x^\a(t)} \right].
\label{ansatz.singular.theta}
\end{equation}
From Eq.~(\ref{ansatz.singular.theta}) we can calculate the Fourier transforms as
\begin{equation}
(\partial_i \thv)({\bf q}, t) = i \frac{h}{m} \epsilon_{ik} \frac{q_k}{q^2} 
\sum_\a e^\a e^{-i {\bf q}\cdot {\bf x}^\a(t)},
\label{gradient.theta.v}
\end{equation}
\begin{equation}
(\partial_t \thv)({\bf q}, t) = - i \frac{h}{m} \epsilon_{ik} \frac{q_k}{q^2} 
\sum_\a e^\a s^\a_i(t) e^{-i {\bf q}\cdot {\bf x}^\a(t)},
\label{partial.tau.theta.v}
\end{equation}
where ${\bf s}^\a (t) = d{\bf x}^\a(t)/dt$ is the velocity of the $\alpha$-th vortex.
Substituting Eqs.~(\ref{gradient.theta.v}) and (\ref{partial.tau.theta.v}) into Eq.~(\ref{SS.action.V2}), $S_2^\text{V}$ becomes
\begin{widetext}
\begin{eqnarray}
S_2^\text{V} &=&
- \frac{\rho_\text{s}\hbar^2}{2m^2} \sum_{\a, \b} e^\a e^\b \int d t \;
F_0({\bf x}^\a(t) - {\bf x}^\b(t))
\nonumber
\\
&&
- \frac{\hbar^2 \tilde{\Gamma}^2}{8m^2\rn}  
\sum_{\a, \b} e^\a e^\b \int d t \int d t^\prime \;
G_1({\bf x}^\a(t) - {\bf x}^\b(t^\prime),t-t^\prime;c_T),
\nonumber
\\
&&
- \frac{\hbar^2}{2m^2} \sum_{\a, \b} e^\a e^\b \int d t \int d t^\prime
s_i^\a (t) s_j^\b (t^\prime) \left( \delta_{ij} \p_k^\a \p_k^\b - \p_i^\a \p_j^\b
\right)
G_2({\bf x}^\a(t) - {\bf x}^\b(t^\prime),t-t^\prime;c_L, c_2)
\label{Vortex.action.Appendix.F}
\end{eqnarray}
\end{widetext}
where $\p_i^\a \equiv \p / \p x_i^\a(t)$ and $\p_i^\b \equiv \p / \p x_i^\b(t^\prime)$, and we defined the auxiliary integrals
\begin{equation}
F_0({\bf x}) =
\int d^2 q \frac{1}{q^2} e^{-i{\bf q} \cdot {\bf x}},
\label{integral.1.appendix.F}
\end{equation}
\begin{equation}
G_1({\bf x},t;c) = \int d^2 q \int \frac{d\omega}{2\pi} \frac{1}{q^2} 
\frac{\omega^2}{\omega^2 - c^2 q^2} e^{-i{\bf q} \cdot {\bf x} + i\omega t},
\label{integral.2.appendix.F}
\end{equation}
\begin{equation}
G_2({\bf x},t;c,\tilde{c}) = \int d^2 q \int \frac{d\omega}{2\pi} \frac{1}{q^2} 
\frac{ a \omega^2 - b q^2}{\Delta}
e^{-i {\bf q} \cdot {\bf x} + i\omega t},
\label{integral.3.appendix.F}
\end{equation}
with $c$ and $\tilde{c}$ in $G_1$ and $G_2$ the sound speeds ($c_L$, $c_T$, and $c_2$). 
In Eq.~(\ref{integral.3.appendix.F}), $\Delta$ is defined in Eq.~(\ref{Delta.determinant}), the coefficients $a$ and $b$ are
\begin{equation}
a= \rn \rs + \frac{\tilde{\Gamma}^2}{4},
\label{coefficient.a.G2}
\end{equation}
and 
\begin{equation}
b=\rs \left( \tilde{\Lambda} + 2\tilde{\mu} + \frac{\tilde{\Gamma}^2}{4 \tilde{\chi}} \right).
\label{coefficient.b.G2}
\end{equation}
Therefore, the action for vortices can be derived by evaluating the auxiliary integrals defined in Eqs.~(\ref{integral.1.appendix.F})-(\ref{integral.3.appendix.F}). 

Before simplifying the action, Eq.~(\ref{Vortex.action.Appendix.F}), it is worth stepping back and considering the special case of a \textit{static} configuration of vortices. In the static limit, $\textbf{x}^\alpha$ is independent of the time $t$, and the velocity $\textbf{s}^\alpha = 0$, so the third term of Eq.~(\ref{Vortex.action.Appendix.F}) is zero. For the second term, we introduce new time variables, $\tau = t-t^\prime$ and $\bar{t} = (t+t^\prime)/2$; integrating $G_1$ over the difference variable $\tau$ gives $2\pi \delta (\omega)$, and performing the $\omega$ integral in $G_1$ renders the second term zero as well. Finally, if we introduce a momentum cutoff $q_0$ into the expression for $F_0$, we obtain
\begin{equation}
	F_0 ({\bf x}) = \int d^2q\, \frac{e^{-i \bf{q}\cdot \bf{x}}}{q^2+q_0^2}  = 2\pi K_0(q_0x),
\end{equation}
where $K_0(z)$ is the modified Bessel function of the second kind. For small argument $K_0(z)\sim -\ln (z/2) - \gamma_E$, with $\gamma_E = 0.5772...$ Euler's constant, and up to a constant we obtain the static action
\begin{equation}
	S_2^{\text{static}} =  \frac{2\pi \rho_\text{s}\hbar^2}{2m^2} \int dt \sum_{\a, \b} e^\a e^\b \ln (|{\bf x}^\a - {\bf x}^\b|/\xi). 
\label{static.action}
\end{equation}
with $\xi$ a cutoff. This is exactly the interaction energy of a collection of $n_V$ point vortices in two dimensions \cite{Donnelly1991a}.

Returning to the full action, Eq.~(\ref{Vortex.action.Appendix.F}), we evaluate $G_1$ by separating out the local term, resulting in
\begin{eqnarray}
G_1({\bf x},t;c) &=& 
\int d^2 q \int \frac{d\omega}{2\pi} \frac{1}{q^2} 
\left( 1 - \frac{c^2 q^2}{\omega^2 - c^2 q^2} \right) e^{-i{\bf q} \cdot {\bf x} + i\omega t}
\nonumber
\\
&=& \d(t) F_0({\bf x})
+ c^2 F_1({\bf x},t;c),
\label{function.G1}
\end{eqnarray}
where 
\begin{equation}
F_1({\bf x},t;c) = 
\int d^2 q \int \frac{d\omega}{2\pi}  
\frac{1}{\omega^2 - c^2 q^2} e^{-i{\bf q} \cdot {\bf x} + i\omega t} .
\label{F1}
\end{equation}

Next, we separate $G_2$ into two terms by expanding in partial fractions:
\begin{equation}
G_2({\bf x},t;c,\tilde{c}) =
\frac{A(c,\tilde{c})}{\rn} F_2({\bf x},t;c) + \frac{A(\tilde{c},c)}{\rn} F_2({\bf x},t;\tilde{c}),
\label{function.G2}
\end{equation}
where
\begin{equation}
F_2({\bf x},t;c) =  
\int d^2 q \int \frac{d\omega}{2\pi} \frac{1}{q^2}
\frac{1}{\omega^2 - c^2 q^2} e^{-i{\bf q} \cdot {\bf x} + i\omega t},
\end{equation}
\begin{equation}
A(c,\tilde{c}) \equiv \frac{a c^2-b}{c^2-\tilde{c}^2},
\label{coefficient.A.of.G2}
\end{equation}
with $a$ and $b$ given by Eqs.~(\ref{coefficient.a.G2})-(\ref{coefficient.b.G2}). 
The second term in Eq.~(\ref{Vortex.action.Appendix.F}) contains spatial derivatives of $G_2$. The first spatial derivatives equal minus the Laplacian of $G_2$ because $\p_i^\b = - \p_i^\a$. Next, following Ref.~\cite{Eckern1989a}, the second spatial derivatives with vortex velocities can be converted into temporal derivatives as $s_i^\a (t) s_j^\b (t^\prime) \p_i^\a \p_j^\b = \p_t \p_{t^\prime}$.
Hence, we can evaluate the second and third terms with $G_2$ in Eq.~(\ref{Vortex.action.Appendix.F}) by calculating $\p_t^2 F_2$ and $\p^2 F_2$ instead. In fact, they can be written in terms of $F_0$ and $F_1$:
\begin{equation}
\p^2 F_2({\bf x},t;c) = -F_1({\bf x},t;c),
\label{Laplacian.F2}
\end{equation}
and
\begin{equation}
\p_t^2 F_2({\bf x},t;c) = -F_0({\bf x}) \delta(t) - c^2 F_1({\bf x},t;c).
\label{partial.tau.F2}
\end{equation}

Finally, we can combine the local contributions of $G_1$ and $\p_t^2 G_2$ with the first term of $S_2$, resulting in
\begin{widetext}
\begin{equation}
\text{Local terms in $S_2$} = 
- \frac{\hbar^2}{2m^2} \left( \rs + \frac{\tilde{\Gamma}^2}{4 \rn} - \frac{a}{\rn} \right)
\sum_{\a, \b} e^\a e^\b \int d t 
F_0({\bf x}^\a(t) - {\bf x}^\b(t)) .
\end{equation}
Note that the above sum of local terms vanishes as $a = \rs \rn + \tilde{\Gamma}^2 /4$. After the cancellation of the local term, with the non-local terms from Eqs.~(\ref{F1}), (\ref{Laplacian.F2}) and (\ref{partial.tau.F2}), we finally obtain the action for vortices in terms of their positions and velocities:
\begin{eqnarray}
S_2^\text{V} &=&
- \frac{\hbar^2 \tilde{\Gamma}^2 c_T^2}{8 m^2 \rho_\text{n} } 
\sum_{\a , \b} e^\a e^\b 
\int dt \int dt^\prime
F_1 ({\bf x^\a}(t) - {\bf x^\b}(t^\prime), t - t^\prime, c_T)
\nonumber
\\
&&
- \frac{\hbar^2 A(c_L,c_2)}{2 m^2 \rho_\text{n}} 
\sum_{\a ,\b} e^\a e^\b 
\int dt \int dt^\prime
\left[ s^\a_i(t) s^\b_i(t^\prime) - c_L^2 \right]
F_1 ({\bf x^\a}(t) - {\bf x^\b}(t^\prime), t - t^\prime, c_L)
\nonumber
\\
&&
- \frac{\hbar^2 A(c_2, c_L)}{2 m^2 \rho_\text{n}} 
\sum_{\a ,\b} e^\a e^\b 
\int dt \int dt^\prime
\left[ s^\a_i(t) s^\b_i(t^\prime) - c_2^2 \right]
F_1 ({\bf x^\a}(t) - {\bf x^\b}(t^\prime), t - t^\prime, c_2) .
\label{SS.action.4}
\end{eqnarray}
\end{widetext}
This action is one of the main results of this work.  Similar to Refs.~\cite{Eckern1989a, Arovas1997a}, vortex interactions are mediated by sound modes, generating effective non-local ``Coulomb'' potential terms. In a two-dimensional supersolid, three sound modes are present: a longitudinal second sound mode from the superfluid, and longitudinal and transverse sound modes associated with the elasticity of the solid. In the absence of coupling between the superfluidity and the elasticity, $\tilde{\Gamma}=0$ $(\rho_\text{n} = \rho^2\chi\gamma)$, we obtain $A(c_L,c_2)=0$ and $A(c_2,c_L)=\rho_\text{n}\rho_\text{s}$, and the above action for vortices reduces to Eq.~(3.25) of Eckern and Schmid~\cite{Eckern1989a}. It is worth noting that the velocity-velocity interactions only occur through longitudinal modes--there is not a coupling induced by the transverse elastic mode. Finally, if we take the static limit of Eq.~(\ref{SS.action.4}), we obtain the static action in Eq.~(\ref{static.action}).

\subsection{Inertial Mass of a Vortex}

In a compressible medium (a fluid or a solid) there is a retarded self-interaction of a vortex that yields a frequency-dependent inertial mass \cite{Baym1983a,Eckern1989a,Eckern1993a,Fazio1994a,Sonin1998a,Chudnovsky2003a,Arovas1997a,Simula2018a}. 
For a single vortex of charge $e^\alpha$, the action above reduces to
\begin{widetext}
	
\begin{eqnarray}
S_2^\text{V} &=&
- \frac{\hbar^2 \tilde{\Gamma}^2 c_T^2 (e^{\alpha})^2}{8 m^2 \rho_\text{n}} 
\int dt \int dt^\prime
F_1 ({\bf x}(t) - {\bf x}(t^\prime), t - t^\prime, c_T)
\nonumber
\\
&&
- \frac{\hbar^2 (e^{\alpha})^2 A(c_L,c_2)}{2 m^2 \rho_\text{n}} 
\int dt \int dt^\prime
\left[ s_i(t) s_i(t^\prime) - c_L^2 \right]
F_1 ({\bf x}(t) - {\bf x}(t^\prime), t - t^\prime, c_L)
\nonumber
\\
&&
- \frac{\hbar^2 (e^{\alpha})^2 A(c_2, c_L)}{2 m^2 \rho_\text{n}} 
\int dt \int dt^\prime
\left[ s_i(t) s_i(t^\prime) - c_2^2 \right]
F_1 ({\bf x}(t) - {\bf x}(t^\prime), t - t^\prime, c_2) .
\label{SS.action.single}
\end{eqnarray}

\end{widetext}
The effective inertial mass of a vortex can be determined by finding the coefficient of $s_i(t) s_i(t^\prime)$ in the low velocity limit, $c^2(t - t^\prime)^2 \gg |{\bf x}(t) - {\bf x}(t^\prime)|^2$ for all three sound speeds. In this limit, the auxiliary function $F_1$, Eq.~(\ref{F1}), can be expanded as 
\begin{equation}
F_1({\bf x},t;c) \simeq - \frac{\pi}{c^2 |t|} + \frac{\pi |{\bf x}|^2}{2 c^4 |t|^3}.
\end{equation} 
Note that the second term in the expansion is also in the form of a kinetic energy term. The first term, on the other hand, does not contribute to the inertial mass but to the rest mass. Then, the action for a single vortex can be separated into two parts
\begin{equation}
S^\text{V}_2=S^\text{V}_\text{rest}+S^\text{V}_\text{massive}.
\end{equation}
The static action $S^\text{V}_\text{rest}$ has a logarithmic divergence for both large and small $t$ that arises when the first term of $F_1$ is integrated over $t$ as found by Arovas and Freire \cite{Arovas1997a}. As they noted, the small $t$ divergence can be regularized with a cut-off associated with the core size of the vortex, whereas the large $t$ divergence is inevitable due to the logarithmic divergence of the energy of a static vortex. On the other hand, the massive action $S^\text{V}_\text{massive}$ is in a form
\begin{equation}
S^\text{V}_\text{massive} = \frac{1}{2} \int \frac{d\omega}{2\pi} M^\text{V}(\omega) \omega^2 x_i (\omega) x_i (-\omega),
\end{equation}
where the frequency-dependent mass of a vortex in a supersolid has three contributions:
\begin{equation}
M^\text{V}(\omega) = M_{c_T}^\text{V}(\omega) + M_{c_L}^\text{V}(\omega) + M_{c_2}^\text{V}(\omega).
\end{equation}
When the expansion of $F_1$ is used in Eq.~(\ref{SS.action.single}), the massive action $S^\text{V}_\text{massive}$ has two terms at second order of the vortex velocity over the sound speed:
\begin{equation}
I_1 = \int dt \int dt^\prime \frac{s_i(t) s_i(t^\prime)}{|t - t^\prime|} ,
\end{equation}
and
\begin{equation}
I_2 = \int dt \int dt^\prime \frac{|{\bf x}(t) - {\bf x}(t^\prime)|^2}{|t - t^\prime|^3} .
\end{equation}
We then calculate the Fourier transforms of the integrands of $I_1$ and $I_2$ and find
\begin{equation} 
I_1 = -2 \int \frac{d\omega}{2\pi} \omega^2 \text{ci}(\epsilon |\omega|) x_i (\omega) x_i (-\omega),
\end{equation}
and
\begin{widetext}
\begin{equation}
I_2 = -2 \int \frac{d\omega}{2\pi} \omega^2 \left[  \text{ci}(\epsilon |\omega|) 
- \frac{1 - \cos(\epsilon \omega)}{\epsilon^2 \omega^2} - 
    \frac{\sin(\epsilon \omega)}{\epsilon \omega} \right] x_i (\omega) x_i (-\omega),
\end{equation}
where $\text{ci}(x)$ is the cosine integral, 
\begin{equation}
	\text{ci} (x) = -\int_x^\infty \frac{\cos{t}}{t}\, dt ,
\end{equation}
and $\epsilon$ is a cut-off introduced to regulate the infrared divergence in temporal integrals. Again, this low-frequency cut-off is associated with the core structure of a vortex. 
With the above $I_1$ and $I_2$, each contribution to the mass term can be found:  
\begin{equation}
M_{c_T}^\text{V}(\omega) =
- \frac{\pi \hbar^2 (e^{\alpha})^2 \tilde{\Gamma}^2}{4 m^2 \rho_\text{n} c_T^2} 
\Bigg[
\text{ci}(\epsilon |\omega|) - \frac{1-\cos(\epsilon \omega)}{\epsilon^2 \omega^2} 
- \frac{\sin(\epsilon \omega)}{\epsilon \omega}
\Bigg],
\end{equation}
\begin{equation}
M_{c_L}^\text{V}(\omega) =
- \frac{\pi \hbar^2 (e^{\alpha})^2 A(c_L,c_2)}{2 m^2 \rho_\text{n} c_{L}^2} 
\Bigg[
\text{ci}(\epsilon |\omega|) + \frac{1-\cos(\epsilon \omega)}{\epsilon^2 \omega^2} 
+ \frac{\sin(\epsilon \omega)}{\epsilon \omega}
\Bigg],
\end{equation}
\begin{equation}
M_{c_2}^\text{V}(\omega) =
- \frac{\pi \hbar^2 (e^{\alpha})^2 A(c_2,c_L)}{2 m^2 \rho_\text{n} c_{2}^2} 
\Bigg[
\text{ci}(\epsilon |\omega|) + \frac{1-\cos(\epsilon \omega)}{\epsilon^2 \omega^2} 
+ \frac{\sin(\epsilon \omega)}{\epsilon \omega}
\Bigg].
\end{equation}
Note that the signs of the second and third terms in the square bracket of $M_{c_T}^\text{V}$ are not errors. The additional contribution from the square of vortex velocity present in the second and third terms in Eq.~(\ref{SS.action.single}) makes the sign change for similar terms in $M_{c_L}^\text{V}$ and $M_{c_2}^\text{V}$. 
It is important to note that without the coupling of superfluid to elasticity [$\tilde{\Gamma}=0$ $(\rho_\text{n} = \rho^2\chi\gamma)$, $A(c_L,c_2)=0$ and $A(c_2,c_L)=\rho_\text{n}\rho_\text{s}$], $M_{c_T}^\text{V}$ and $M_{c_L}^\text{V}$ vanish, and we recover the vortex mass $M_{c_2}^\text{V}$ found by Arovas and Freire \cite{Arovas1997a}.

In the low frequency limit, $[1-\cos(\epsilon\omega)]/\epsilon^2 \omega^2 \simeq 1/2$, $\sin(\epsilon\omega)/\epsilon\omega \simeq 1$, and $\text{ci}(\epsilon |\omega|) \simeq \gamma_E + \ln(\epsilon |\omega|)$, and we find the logarithmic frequency-dependent vortex mass, 
\begin{equation}
M^\text{V}(\omega) = - \frac{\pi \hbar^2 (e^{\alpha})^2}{m^2} \Bigg[ \rho^2\chi 
+ \frac{1}{4}\tilde{\Gamma}^2 
\left( \frac{1}{\tilde{\mu}} + \frac{1}{\tilde{\Lambda} + 2\tilde{\mu}} \right) \Bigg] \left[ \gamma_E + \frac{3}{2} +  \ln(|\omega| \epsilon)\right]
+ \frac{3 \pi \hbar^2 (e^{\alpha})^2 \tilde{\Gamma}^2}{4 m^2 \rho_\text{n} c_T^2} .
\label{vortex.mass}
\end{equation}
\end{widetext}
If $\tilde{\Gamma}=0$, the above frequency-dependent mass of a vortex in superfluid becomes the mass obtained by Eckern and Schmid \cite{Eckern1989a}. However, in their work this vortex mass was interpreted as an effective viscosity.

\subsection{Magnus Force}

As shown in Refs.~\cite{Ao1993a,Zhu1996a,Thouless1996a}, $S_1$ produces a transverse Magnus force on vortex. Since $\partial_t = \sum_{\a} s_i^\a(t) \partial / \partial x_i^\a(t)$, 
$S_1^\text{V}$ can be rewritten in terms of the vortex velocity and an effective vector potential,
\begin{equation}
S_{1}^\text{V} = \sum_{\a} \int dt A_i^\a[{\bf x}^\a(t)] s_i^\a(t),
\end{equation}
where we have defined 
\begin{equation}
A_i^\a({\bf x}^\a) = - \rho \int d^2 x \frac{\partial \theta^\a( {\bf x} - {\bf x}^\a)}{\partial x_i^\a}.
\end{equation}
Therefore, the $\a$-th vortex moves with a velocity ${\bf s}^\a(t)$ in an effective magnetic field given by $\nab_{{\bf x}^a} \times {\bf A}^\a({\bf x}^\a) = - h \rho e^\a \hat{z} / m$. The Magnus force \cite{Ao1993a,Zhu1996a,Thouless1996a} acting on the $\a$-th vortex is then given by
\begin{equation}
{\bf F}_\text{Magnus}^\a = - \frac{h \rho e^\a}{m} {\bf s}^\a(t) \times \hat{z},
\end{equation}
which is a force perpendicular to the vortex velocity, analogous to the Lorentz force on a charged particle in an applied magnetic field. 

\section{Summary}

We have investigated the dynamics of vortices in supersolids starting from an effective action for a supersolid. After separating the superfluid phase into smooth and singular components, we eliminated the smooth fields, resulting in an effective action for vortex degrees of freedom. This action consists of retarded, nonlocal interactions in spacetime, mediated by the gapless propagating modes in the supersolid: a longitudinal ``second sound'' mode (with speed $c_2$), a longitudinal elastic mode (with speed $c_L$), and a transverse elastic mode (with speed $c_T$). In the static limit, the action reduces to the usual logarithmic interaction between point vortices in superfluid films; we also recover the Magnus force on the vortices from the topological ``Berry phase'' term in the action. Finally, when retardation effects are included, the vortices acquire a frequency-dependent inertial mass due to the coupling to the gapless modes. 

\begin{acknowledgments}
	
This work was started some fifteen years ago, when the authors were at the University of Florida. We thank our UF colleagues Mark Meisel, Yoonseok Lee, Kinjal Dasbiswas, and Debajit Goswami for helpful discussions and collaboration. ATD would like to thank Leo Radzihovsky and Paul Goldbart for helpful discussions; and Wilhelm Zwerger and Johannes Hofmann for drawing our attention to the interesting physics of dipolar supersolids. This work was performed in part at the Aspen Center for Physics, which is supported by National Science Foundation grant PHY-2210452.

\end{acknowledgments}

\appendix

\section{Supersolid Lagrangian}

In this appendix we give a brief summary of our prior work, which derived an effective Lagrangian for a supersolid using variational principles developed for other hydrodynamic systems \cite{Yoo2010a,YooThesis2009}. We start with the action for a supersolid,
\begin{equation}
S_\text{SS} = \int dt\int d^3x \lag_\text{SS},
\label{Eulerian.Action}
\end{equation}
where $\lag_\text{SS}$ is the Lagrangian density. In the Eulerian specification (in which all fields are depicted at positions $\mathbf{x}$ and time $t$), the Lagrangian consists of the kinetic energy densities due to the super- and normal components, and the internal energy density:
\begin{eqnarray}
\lag_\text{SS} &=& \half \rs_{ij} \vs_i \vs_j 
+ \half \rn_{ij}\vn_i \vn_j \nonumber
\\
&& - U_\text{SS}(\rho,\rs_{ij},s,R_{ij}), \label{lagrangian1}
\end{eqnarray}
where $\rs_{ij}$ and $\rn_{ij}$ are the superfluid and normal density tensors with total density $\rho$ satisfying $\rho\,\delta_{ij}=\rs_{ij}+\rn_{ij}$, and where $\bvs$ and $\bvn$ are the velocities of the super- and normal components. The internal energy density $U_\text{SS}$ is taken to be dependent on $\rho$, $\rs_{ij}$, the entropy density $s$, and the deformation tensor 
\begin{equation}
R_{ij} \equiv \p_i R_j,
\label{deform.tensor}
\end{equation}
with ${\bf R}$ the coordinate affixed to material elements.
For a solid (either a normal solid or supersolid) the internal energy depends on the deformation tensor to account for the broken translational symmetry. As shown in Ref.~\cite{Yoo2010a}, the internal energy density satisfies the thermodynamic relation
\begin{eqnarray}
d U_\text{SS} &=& T ds + \left[ \mu + \half (\vn_i - \vs_i)^2
\right] d \rho - \lambda_{ik} d R_{ik} \nonumber
\\
&& - \half (\vn_i - \vs_i)(\vn_j - \vs_j) d{\rho_s}_{ij},
\label{thermo.Uss}
\end{eqnarray}
where $T$ is the temperature, $\mu$ the chemical potential per unit mass, and $\lambda_{ij}$ the stress tensor. 

The next step is to incorporate the constraints that result from conservation laws and broken symmetries. In three dimensions a fluid has five conserved quantities (mass, entropy, and three components of momentum); a solid has an additional three broken translational symmetries; and a supersolid also has a broken gauge symmetry.   The conservation laws for a supersolid are the continuity equation for mass
\begin{equation}
\p_t \rho + \p_i j_i=0, \label{SS.continuity}
\end{equation}
where the mass current $j_i$ is
\begin{equation}
j_i =\rs_{ij} \vs_j + (\rho \d_{ij} - \rs_{ij}) \vn_j,
\label{mass.current}
\end{equation}
and the entropy conservation law 
\begin{equation}
\p_t s + \p_i (s \vn_i) = 0, \label{SS.entropy}
\end{equation}
which only involves $\bvn$ since the entropy is transported by the normal fluid. To account for the broken translational symmetry, Lin's constraint \cite{Lin1963a} is used:
\begin{equation}
\frac{D_n R_i}{D t} =0, \label{Lin}
\end{equation}
where $D_n / D t \equiv \p_t + \vn_i \p_i$. After incorporating all of the constraints, Eqs.~(\ref{SS.continuity})-(\ref{Lin}), into the Lagrangian density Eq.~(\ref{lagrangian1}) using the Lagrange multipliers (or ``Clebsch potentials'') $\alpha$, $\phi$, and $\beta_i$, we obtain the supersolid Lagrangian
\begin{eqnarray}
\lag_\text{SS} &=& \half \rs_{ij} \vs_i \vs_j + \half (\rho \d_{ij}
-\rs_{ij})\vn_i \vn_j \nonumber
\\
&& - U_\text{SS}(\rho,\rs_{ij},s,R_{ij}) + \alpha \bigg[ \p_t s +
\p_i (s \vn_i) \bigg] \nonumber
\\
&& + \phi \Bigg\{\p_t \rho + \p_i \bigg[ \rs_{ij} \vs_j + (\rho
\d_{ij} - \rs_{ij}) \vn_j \bigg]\Bigg\} \nonumber
\\
&& + \beta_i \bigg[ \p_t(sR_i) + \p_j (s R_i \vn_j) \bigg].
\label{lag2}
\end{eqnarray}

The hydrodynamic equations for supersolids are obtained by taking variations of the action Eq.~(\ref{Eulerian.Action}) with the Lagrangian Eq.~(\ref{lag2}). While the variations with respect to the Lagrange multipliers $\alpha$, $\beta_i$, and $\phi$ reproduce the imposed constraints, Eqs.~(\ref{SS.continuity})-(\ref{Lin}), the variations with respect other dynamical variables $\rho$, $\rs_{ij}$, $s$, $\vn_i$, $\vs_i$, and $R_i$ yield the following hydrodynamic equations:\\
i) the superfluid velocity as a potential flow
\begin{equation}
\vs_i = \p_i \phi; \label{SS.velocity}
\end{equation}
ii) the Josephson equation
\begin{equation}
\p_t \vs_i = - \p_i \mu - \half \p_i \vs^2; \label{josephson.eqn}
\end{equation}
iii) the momentum conservation law equation 
\begin{equation}
\partial_t j_i + \partial_j \Pi_{ij} = 0,
\label{momentum.conservation.eqn}
\end{equation}
where $j_i$ is the mass current given in Eq.~(\ref{mass.current}), and $\Pi_{ij}$ is the non-dissipative stress tensor
\begin{eqnarray}
\Pi_{ij} &=& \rho\vs_i\vs_j + \vs_i p_j + \vn_j p_i -
R_{ik}\lambda_{jk} \nonumber
\\
&& -\bigg[\epsilon - Ts - \mu \rho - (\vn_j - \vs_j) p_j
\bigg]\delta_{ij}, \label{stress.tensor}
\end{eqnarray}
where $p_i \equiv (\rho \d_{ij} - \rs_{ij}) (\vn_j - \vs_j)$ and
$\epsilon$ is the energy density measured in the frame where the super-components are stationary. 

In Ref.~\cite{Yoo2010a}, we showed that the above hydrodynamic equations derived from the Lagrangian Eq.~(\ref{lag2}) are equivalent to the non-dissipative supersolid hydrodynamics
developed by Andreev and Lifshitz \cite{Andreev1969a}, Saslow \cite{Saslow1977a}, and Liu \cite{Liu1978a}. Furthermore, we found that as a solid with point defects undergoes a supersolid transition, the gauge symmetry is broken and a pair of longitudinal second sound modes appear.

When the Lagrange multipliers $\alpha$ and $\beta_i$ are eliminated by integrating the action by parts, after neglecting boundary terms we obtain a more familiar form of the Lagrangian: 
\begin{eqnarray}
\lag_\text{SS} &=& -\rho \p_t \phi - \half \rs_{ij} \p_i \phi \p_j
\phi + \half (\rho \d_{ij} -\rs_{ij})\vn_i \vn_j \nonumber
\\
&& - (\rho \d_{ij} -\rs_{ij}) \vn_j \p_i \phi -
f(\rho,\rs_{ij},T,R_{ij}), \label{lag4}
\end{eqnarray}
where $f \equiv U_\text{SS} - Ts$ satisfies the thermodynamic relation
\begin{eqnarray}
d f &=& - s dT + \left[ \mu + \half (\vn_i - \vs_i)^2 \right] d \rho
- \lambda_{ik} d R_{ik} \nonumber
\\
&& - \half (\vn_i - \vs_i)(\vn_j - \vs_j) d\rs_{ij}.
\label{thermodynamic.relation.f}
\end{eqnarray}
The reader can find a valuable discussion in Ref.~\cite{Yoo2010a} on the relation between this Lagrangian and those derived in Refs.~\cite{Son2005a, Josserand2007a}.

The quadratic action Eq.~(\ref{SS.action1}) is derived by expanding the Lagrangian Eq.~(\ref{lag4}) up to second order in small fluctuations $\d \rho$, $\d \rs_{ij}$, $\p_i \phi$, $\p_t \phi$, and $w_{ij} \equiv \p_i u_j$ with $u_i$ being lattice fluctuations away from equilibrium, resulting in
\begin{eqnarray}
\lag_\text{SS}^\text{quad} &=& - \rho_0 \p_t \phi - {\lambda_0}_{ij}
w_{ij} - \mu_0 \d \rho - \d \rho \p_t \phi \nonumber
\\
&& - \half \rho_0 (\p_i \phi)^2 - \frac{\p \mu}{\p
w_{ij}}\bigg|_\rho \d \rho \, w_{ij} - \half \frac{\p \mu}{\p
\rho}\bigg|_{w_{ij}} (\d \rho)^2 \nonumber
\\
&& + \half {\rn_0}_{ij} \left( \p_t u_i - \p_i \phi \right) 
\left(\p_t u_j - \p_j \phi \right) \nonumber
\\
&& - \half \frac{\p \lambda_{ij}}{\p w_{lk}}\bigg|_\rho w_{ij}
w_{lk}, \label{SS.quad.lag}
\end{eqnarray}
where the subscript `0' refers to equilibrium quantities. If the density fluctuation $\delta\rho$ is integrated out and $\theta = \phi + \mu_0 t$ is introduced, we obtain the Lagrangian given in Eq.~(\ref{SS.action1}), the starting point of this work. Note that the first two terms in the above Lagrangian are in the form of total derivatives, and \textit{prima facie} would not appear to contribute to the dynamics. However, for singular, multivalued fields that describe vortices or dislocations, these terms are crucial, and need to be retained in the action. 

\bibliography{ref.supersolid.final}

\end{document}